\documentclass{article}

\usepackage{blindtext}
\usepackage[a4paper, total={6in, 8in}]{geometry}
\usepackage[english]{babel}

\usepackage{amsmath}
\usepackage{makecell}
\usepackage{booktabs}
\usepackage{multirow}
\usepackage{float}
\usepackage{graphicx}
 \usepackage{cite}
\usepackage{algorithm}
\usepackage{color}
\providecommand{\keywords}[1]
{
  \small	
  \textbf{\textit{Keywords---}} #1
}

\begin{document}
\title{Acceleration-Based Kalman Tracking for Super-Resolution Ultrasound Imaging \emph{in vivo}}
\author{Biao Huang, Jipeng Yan, Megan Morris, \\
Victoria Sinnett, Navita Somaiah and Meng-Xing Tang,
\thanks{This work was supported by the China Scholarship Council, Chan Zuckerberg Initiative under No. 2020-225443, MRC Confidence in Concept scheme at Imperial College under No. MC-PC-18050, the Engineering and Physical Sciences Research Council under No. EP/T008970/1, the CRUK Convergence Science Centre at The Institute of Cancer Research, London, and Imperial College London (C309/A31546), NHS funding to the National Institute for Health Research (NIHR) Biomedical Research Centre at the Royal Marsden and the Institute of Cancer Research (ICR). KORTUC phase II receives funding from Kortuc Inc., Japan under No. NCT03946202. (Biao Huang and Jipeng Yan contributed equally to this work. Corresponding author: Meng-Xing Tang.}
\thanks{Huang, Yan, Morris and Tang are with Ultrasound Lab for Imaging and Sensing, Department of Bioengineering, Imperial College London, London, UK, SW7 2AZ. (e-mail: b.huang21@imperial.ac.uk; j.yan19@imperial.ac.uk; m.morris20@imperial.ac.uk; mengxing.tang@imperial.ac.uk).}
\thanks{Sinnett is with the Royal Marsden NHS Foundation Trust, London, UK, SW3 6JJ (e-mail: victoria.sinnett@rmh.nhs.uk).}
\thanks{Somaiah  is with the Royal Marsden NHS Foundation Trust and The Institute of Cancer Research, London, UK, SM2 5NG (e-mail: navita.somaiah@icr.ac.uk).}
}
\date{}
\maketitle
\begin{abstract}
    Super-resolution ultrasound can image microvascular structure and flow at sub-wave-diffraction resolution based on localising and tracking microbubbles. Currently, tracking microbubbles accurately under limited imaging frame rates and high microbubble concentrations remains a challenge, especially under the effect of cardiac pulsatility and in highly curved vessels. In this study, an acceleration-incorporated microbubble motion model is introduced into a Kalman tracking framework. The tracking performance was evaluated using simulated microvasculature with different microbubble motion parameters and acquisition frame rates, and \emph{in vivo} human breast tumour ultrasound datasets. The simulation results show that the acceleration-based method outperformed the non-acceleration-based method at different levels of acceleration and acquisition frame rates and achieved significant improvement in true positive rate (up to 10.03\%), false negative rate (up to 28.61\%) and correctly pairing fraction (up to 170.14\%). The proposed method can also reduce errors in vasculature reconstruction via the acceleration-based nonlinear interpolation, compared with linear interpolation (up to 19 $\mu$m). The tracking results from temporally downsampled low frame rate \emph{in vivo} datasets from human breast tumours show that the proposed method has better microbubble tracking performance than the baseline method, if using results from the initial high frame data as reference. Finally, the acceleration estimated from tracking results also provides a spatial speed gradient map that may contain extra valuable diagnostic information.
\end{abstract}

\section{Introduction}
\label{sec:introduction}
Super-resolution ultrasound (SRUS), also known as ultrasound localisation microscopy (ULM), based on localising and tracking sparse microbubbles (MBs), is capable of mapping microvasculature beyond the wave diffraction limit \emph{in vitro}\cite{viessmannAcousticSuperresolutionUltrasound2013,desaillySonoactivatedUltrasoundLocalization2013} and \emph{in vivo}\cite{christensen-jeffriesVivoAcousticSuperResolution2015,erricoUltrafastUltrasoundLocalization2015,lin3DUltrasoundLocalization2017,zhu3DSuperResolutionUS2019,songImprovedSuperResolutionUltrasound2018,andersenSuperResolutionUltrasoundImaging2019,christensen-jeffriesSuperresolutionUltrasoundImaging2020,chenUltrasoundSuperresolutionImaging2020,kanoulasSuperResolutionContrastEnhancedUltrasound2019a}.

Flows in microvasculature can be measured by SRUS via MB tracking, the performance of which is affected by acquisition frame rates and MB concentrations \cite{tangKalmanFilterBasedMicrobubble2020,yanSuperResolutionUltrasoundSparsityBased2022}. As high frame rate acquisitions are not generally available in commercial US system, a low MB concentration is often required to maintain tracking accuracy, leading to long acquisition time to reconstruct vasculature \cite{christensen-jeffriesPoissonStatisticalModel2019}. More tissue motions might happen during a longer acquisition, which makes motion correction more challenging. It is valuable to develop algorithms to track MBs at high concentrations and low frame rates.

Motion models have been used in SRUS to deal with the aforementioned problems. The Kalman filter was combined with the Markov chain Monte Carlo data association algorithm (MCMCDA) \cite{ackermannDetectionTrackingMultiple2016,opacicMotionModelUltrasound2018} or the multiple hypothesis tracking (MHT) procedure to track MBs effectively but with relatively high computational cost \cite{solomonExploitingFlowDynamics2019}. Kalman filtering was also used in \cite{tangKalmanFilterBasedMicrobubble2020} to smooth the MB trajectory after MB pairing. A multi-feature-based tracking algorithm was proposed in \cite{yanSuperResolutionUltrasoundSparsityBased2022}, whereas linear motion model was combined with MB image features in a 2D graph-based assignment framework. A hierarchical algorithm with Kalman filtering has also been developed to track MBs at different speed ranges sequentially \cite{taghaviUltrasoundSuperresolutionImaging2022}. However, all the above-mentioned methods assumed a constant velocity for MBs between adjacent frames, which is not strictly true in vessels with cardiac pulsatility, spatial speed gradient or large vascular curvature. To better estimate MB movement in flow, an unscented Kalman-filtering-based tracking method was proposed by \cite{piepenbrockMicrobubbleTrackingNonlinear2020}, where the motion of an MB between frames was modelled by a curved trajectory with a constant speed.

Movement of MBs are sampled by SRUS at the frame rate of acquisition, which can consequentially generate a discontinuity in estimated speed and/or direction of MB movement between frames in case of significant changes in flow velocity. In general, the problem of discontinuity is exacerbated by lower frame rates and faster flow speeds. After reconstructing super-resolution images, saturation of reconstructed vasculature can be lower at a lower frame rate in the same duration of acquisition. Therefore, interpolating tracked MB locations between frames has been used to fill the gaps and enhance the saturation by linking paired MBs with straight lines \cite{yanSuperResolutionUltrasoundSparsityBased2022} or further adaptively changing distance between interpolated points \cite{tangKalmanFilterBasedMicrobubble2020}. However, the assumption that MBs were moving in straight lines between frames is not true for curved vessels. A nonlinear interpolation for MB trajectory reconstruction is worth exploring to provide more accurate reconstruction of the microvasculature.

MB tracking allows dynamic flow parameters to be mapped at super-resolution, such as flow speed and direction \cite{christensen-jeffriesVivoAcousticSuperResolution2015}, which adds significant value to potential clinical applications of SRUS. Opacic \cite{opacicMotionModelUltrasound2018} and Zhu \cite{zhuSuperResolutionUltrasoundLocalization2022} have shown that regularity of microvascular flow directions can be a potential marker for cancer in human. A previous clinical study of breast cancer found that the acceleration time index is a useful parameter for differentiating benign breast tumours from malignant tumours by using Doppler ultrasonography \cite{mesakiDifferentiationBenignMalignant2003}.

In this study, based on the assumption that MBs may travel in non-straight vessels and may have non-zero acceleration, we aim to improve the MB tracking algorithm by incorporating an acceleration term in the current Kalman filtering framework, to account for changes of flow speed and direction between frames. Curved trajectories of MBs were reconstructed via Kalman state vectors. A spatial speed gradient map calculated from acceleration was presented.
\section{Methods}
\label{sec:methods}
This section firstly describes the acceleration incorporated Kalman tracking framework and a 3D graph-based method for initialising the velocity of new MBs. Next, the nonlinear MB trajectory interpolation method based on estimated acceleration was described. Finally, the proposed method was evaluated on both simulation and \emph{in vivo} datasets.

\subsection{Acceleration-based Kalman Tracking with 3-frame Initialisation}
MBs can be effectively tracked via the graph-based assignment framework \cite{tangKalmanFilterBasedMicrobubble2020}. We have recently developed a framework which pairs MBs between two consecutive frames by minimising the total cost constructed by image features and a linear motion model \cite{yanSuperResolutionUltrasoundSparsityBased2022}. The linear motion model was used to predict the movement of MBs, where each MB is assumed to move at a constant velocity between two adjacent frames. This assumption is not valid when MB moves in a curved vessel and flow acceleration is significant, especially when the acquisition frame rate is low. A more accurate motion model is required.

A linear motion-based Kalman filtering has been applied to MB tracking, where the motion model is incorporated as part of the MB tracking cost. To be more specific, a probability ($p$) that indicates the likelihood of an MB pair between frames is defined as a cost, given below
\begin{equation}Cost_{track}=1/p=1/N(\mu,\Sigma)\label{eq_cost_track}\end{equation}
\begin{equation}\mu = H_{k}S_{k|k-1}\label{eq_mu}\end{equation}
\begin{equation}\Sigma = H_{k}P_{k|k-1}H_{k}^T+R_{k}\label{eq_sigma}\end{equation}
where $N$ is defined by the Gaussian distribution, $H_{k}$ and $R_{k}$ are the observation model and the covariance of observation noise, respectively, $S_{k|k-1}$ and $P_{k|k-1}$ are the predicted state vector and predicted estimate covariance matrix, respectively, and are both predicted by the state transition matrix $F$. The state transition model is shown in Eq. \eqref{eq_s_kk-1} and \eqref{eq_s_matrix}.
\begin{equation}
    S_{k|k-1}= FS_{k-1|k-1}+Q \label{eq_s_kk-1}
\end{equation}

\begin{equation}
    \begin{array}{l}
    \begin{bmatrix}
    x(k)\\
    v_x(k)\\
    a_x(k)\\
    y(k)\\
    v_y(k)\\
    a_y(k)\\ 
    \end{bmatrix}=
    \begin{bmatrix}
    1 & \Delta t & \frac{\Delta t^2}{2} & 0 & 0 & 0\\ 
    0 & 1 & \Delta t & 0 & 0 & 0\\ 
    0 & 0 & 1 & 0 & 0 & 0\\ 
    0 & 0 & 0 & 1 & \Delta t & \frac{\Delta t^2}{2}\\ 
    0 & 0 & 0 & 0 & 1 & \Delta t\\ 
    0 & 0 & 0 & 0 & 0 & 1
    \end{bmatrix}
    \times
    \begin{bmatrix}
    x(k-1)\\
    v_x(k-1)\\
    a_x(k-1)\\
    y(k-1)\\
    v_y(k-1)\\
    a_y(k-1)\\ 
    \end{bmatrix} \\
    +
    \begin{bmatrix}
        \frac{\Delta t^4}{4} &  \frac{\Delta t^3}{2} &  \frac{\Delta t^2}{2} & 0 & 0 & 0 \\ 
    \frac{\Delta t^3}{2} & \Delta t^2 & \Delta t & 0 & 0 & 0\\ 
    \frac{\Delta t^2}{2} & \Delta t & 1 & 0 & 0 &0 \\ 
    0 & 0 & 0 & \frac{\Delta t^4}{4} & \frac{\Delta t^3}{2} & \frac{\Delta t^2}{2}\\ 
    0 & 0 & 0 & \frac{\Delta t^3}{2} & \Delta t^2 & \Delta t\\ 
    0 & 0 & 0 & \frac{\Delta t^2}{2} & \Delta t & 1
    \end{bmatrix} \times
    \sigma^2_{a} \label{eq_s_matrix}
    \end{array}
\end{equation}

In this study, we modelled the MB movement between frames as an accelerated motion. Thus, the state $S$ contains two-dimensional MB location ($x,y$), velocity ($v_{x},v_{y}$) and acceleration ($a_{x},a_{y}$). $\Delta t$ is the time interval between frames. $Q$ is the covariance of processing noise, which is used to describe the uncertainty of the true motion from the motion model. $\sigma^2_{a}$ is the variance of noise when assuming constant acceleration between frames.

\begin{figure}[!t]
    \centerline{\includegraphics[width=6cm]{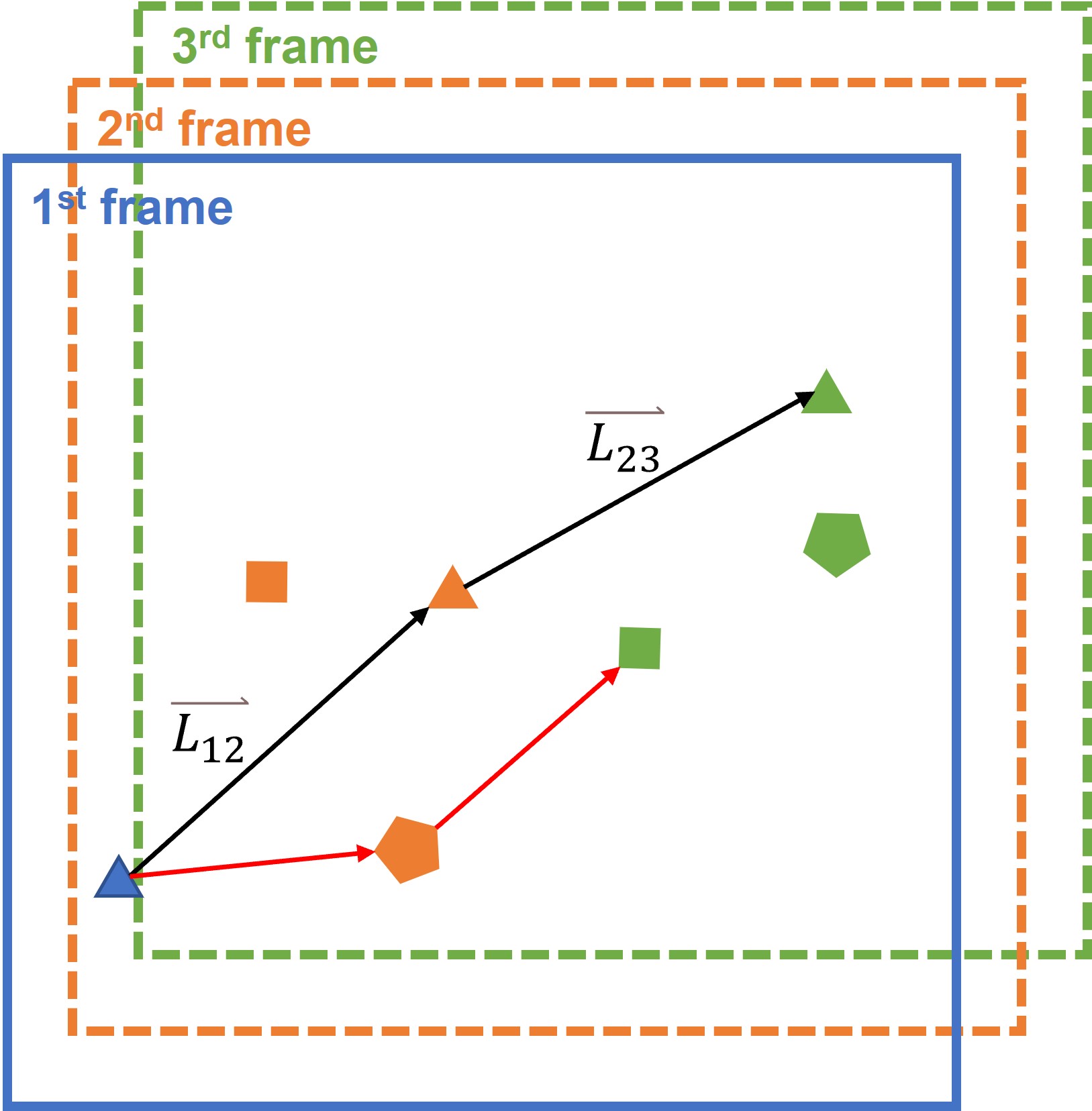}}
    \caption{Illustration of 3-frame initialisation.
    The black line indicates the true MBs pairing. The red line indicates the wrong MBs paring.}
    \label{fig1}
\end{figure}

The state vector of a new MB, including its moving direction, is unknown and can be initiated by using multiple frame tracking. To enable motion model for the new MB, a 3-frame state initialisation method was proposed to give new MBs an initial guess of their states, rather than be set with zero velocity. An assumption for this initialisation is that MB moves smoothly between frames, which indicates the motion direction of MBs between frames will not change dramatically \cite{clarkQuantitativeStudyTrack2019} (Fig. \ref{fig1}). Thus, the true MB pairs can be found by minimising a cost defined by the normalised vector difference among three frames. The cost and state initialisation are given below
\begin{equation}
    Cost_{init}=\frac{\overrightarrow{L_{23}} - \overrightarrow{L_{12}}}{\left \| \overrightarrow{L_{23}} \right \|+\left \| \overrightarrow{L_{12}} \right \|}
    \label{cost_init}
\end{equation}
\begin{equation}
\overrightarrow{v_{init}}=\frac{\overrightarrow{L_{12}} +\overrightarrow{L_{23}}}{2\times\Delta t}
\label{v_init}
\end{equation}
\begin{equation}
\overrightarrow{a_{init}}=\frac{\overrightarrow{L_{23}} - \overrightarrow{L_{12}}}{2\times\Delta t}
\label{a_init}
\end{equation}
where $\overrightarrow{L_{12}}$ and $\overrightarrow{L_{23}}$ are the position vectors between frames, $\overrightarrow{v_{init}}$ and $\overrightarrow{a_{init}}$ indicate the initialised velocity and acceleration vector. The cost was minimised via a 3D graph-based assignment algorithm adapted from a 2D one \cite{sbalzariniFeaturePointTracking2005}, where a topology constraint that each MB can only be paired with no more than one MB at the next frame was set. The motion parameters, including velocity and acceleration, were then initialised from the paired MBs. Only new MBs were paired in the 3D graph-based assignment to initiate their state vectors, and subsequently all MBs in two consecutive frames were paired using the 2D graph-based method.
\subsection{Acceleration-based Nonlinear Interpolation of MB Tracks}
\begin{figure}[!t]
    \centerline{\includegraphics[width=\columnwidth]{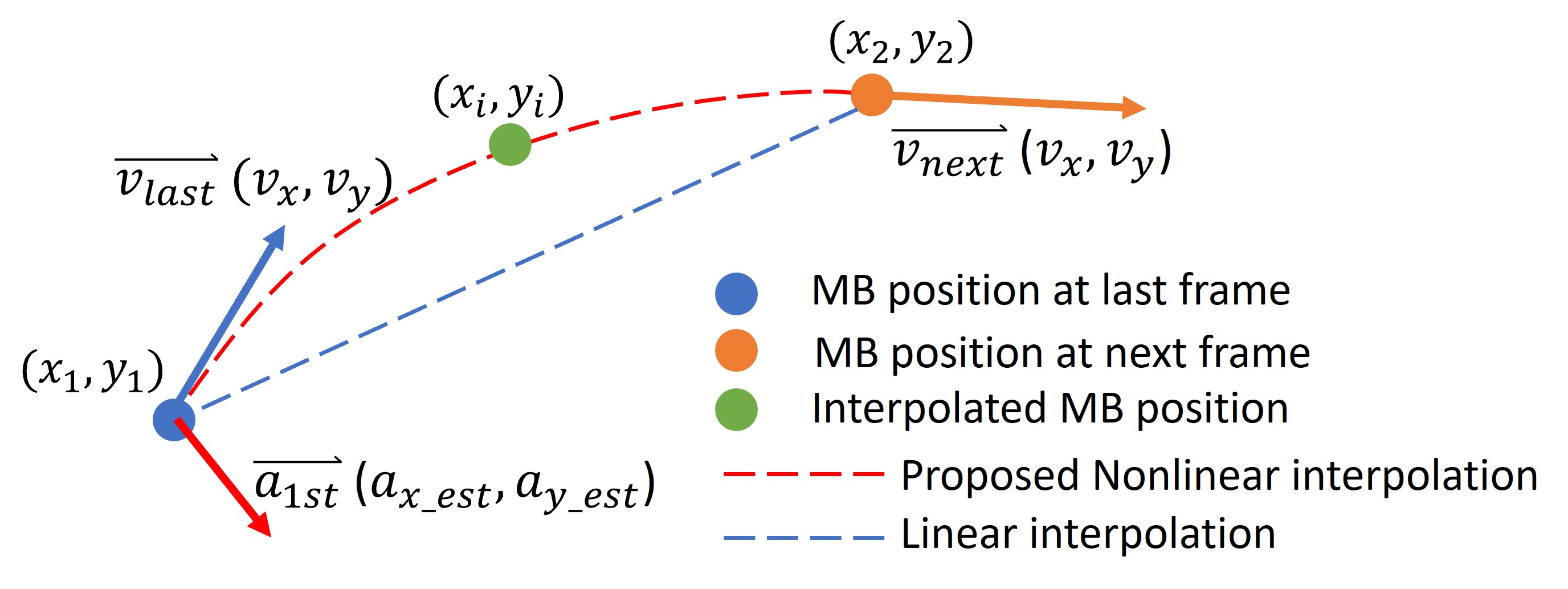}}
    \caption{Examples of microvasculature simulation datasets for interpolation.}
    \label{fig2}
\end{figure}

The microvasculature can be reconstructed by plotting tracked MBs. In this study, based on the estimated motion state from the Kalman-based tracking, we proposed a nonlinear interpolation method for MB trajectory reconstructions (Fig. \ref{fig2}).

The nonlinear interpolation used the MB's acceleration estimated from Kalman states. The missing MB's position between two frames was calculated following the motion model, given below
\begin{equation}
\left\{\begin{matrix}
x_{i} = x_{1} + v_{1x}\times dt+0.5\times a_{x\_est}\times dt^2
\\ 
y_{i} = y_{1} + v_{1y}\times dt+0.5\times a_{y\_est}\times dt^2
\end{matrix}\right.
\label{interpolation}
\end{equation}
where $x_{1}$ and $y_{1}$ indicate the starting coordinates of an MB, $v_{1x}$ and $v_{1y}$ are the estimated MB velocity from Kalman filtering, $a_{x\_est}$ and $a_{y\_est}$ are the estimated MB acceleration for interpolation, and $dt$ is the time interval between the starting position and the position ($x_{i}, y_{i}$) that needs to be interpolated. The acceleration $a_{x\_est}$ and $a_{y\_est}$ are first calculated with Eq. \eqref{interpolation} by replacing ($x_{i}, y_{i}$) with ($x_{2}, y_{2}$), where ($x_{2}, y_{2}$) is the MBs' position at the next frame. The estimated acceleration ($a_{x\_est}$, $a_{y\_est}$) guarantees the continuous trajectory interpolation along all the paired MBs. The speed gradient for each MB can be calculated between the interpolated positions ($x_{i}, y_{i}$) and ($x_{i+1}, y_{i+1}$). The spatial speed gradient was then generated by averaging all the speed gradient at the same positions.

\subsection{Evaluation via Simulations}
The evaluation and comparison of the performance of acceleration-based MB tracking and trajectory interpolation methods are presented in this section. The algorithm and simulation dataset generation were implemented with MATLAB (R2022b, MathWorks, MA, USA).

\subsubsection{Evaluation of MB tracking}
\begin{figure}[!t]
    \centerline{\includegraphics[width=\columnwidth]{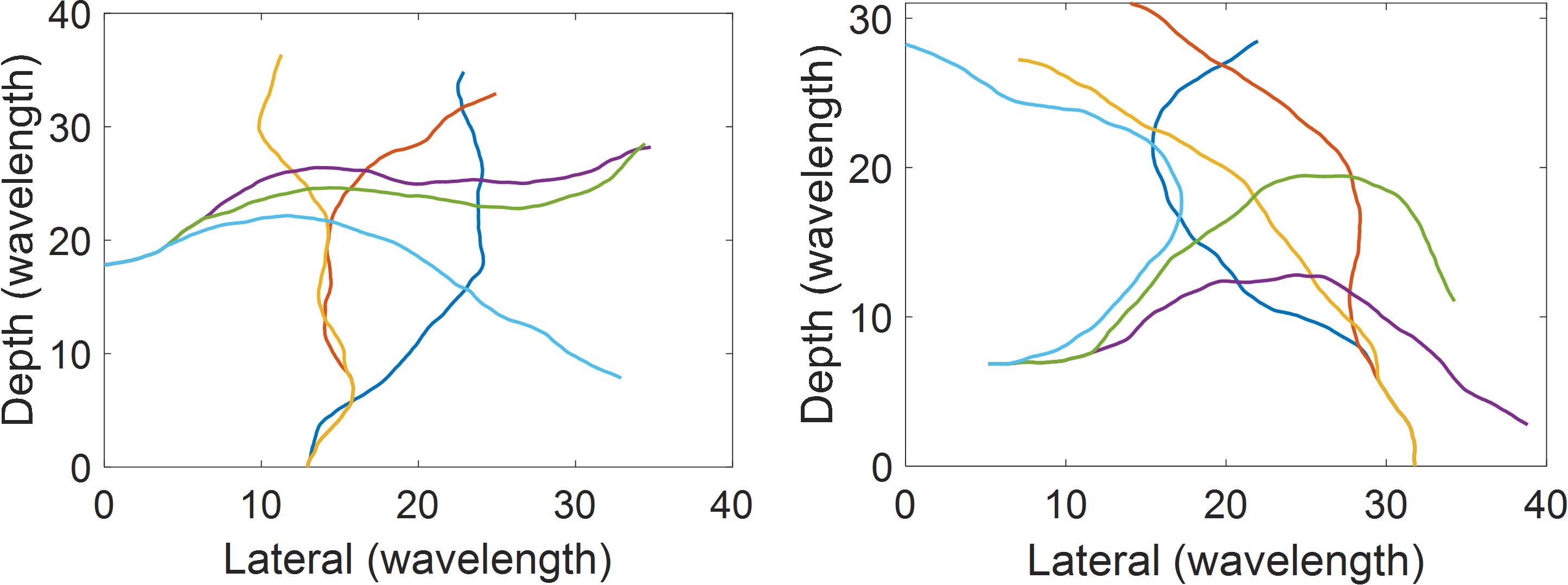}}
    \caption{Examples of microvasculature simulation datasets for tracking.}
    \label{fig3}
\end{figure}

For the MB tracking performance comparison between models with and without acceleration component, we generated six microvasculature simulation datasets. Each dataset had two main vessels each branching into another three downstream vessels (Fig. \ref{fig3}). Three different acquisition frame rates (15 Hz, 25 Hz, and 35 Hz) and three different MB concentrations, estimated from clinical dataset (2.54$\times10^7$ MBs/mL, 3.82$\times10^7$ MBs/mL, 6.36$\times10^7$ MBs/mL), were used in the simulation. To simulate the effect of pulsatile blood flow from cardiac cycles, we accelerated and decelerated the MBs periodically, around the flow speed of 3 mm/s according to a heart rate of 75 bpm, for 30 seconds. Four different flow accelerations were set: 0 mm/s$^2$, 37.5 mm/s$^2$, 75.0 mm/s$^2$, 112.5 mm/s$^2$ \cite{riemerContrastAgentFreeAssessment2022}, and a total of 216 localisation datasets were used to evaluate the tracking performance. 

Both the acceleration-based tracking method (proposed) and non-acceleration-based method (baseline) were tested via a paired \emph{t}-test by SPSS (Version 28.0, IBM Corp, NY, USA) at a significance level of 0.05. Both methods were evaluated among different acceleration and frame rate settings. The metrics for evaluating the tracking performance evaluation were based on \cite{lerendeguiBUbbleFlowField2022} which has been used in a recent super-resolution ultrasound challenge (https://ultra-sr.com) and includes true positive rate, false negative rate and the fraction of correctly paired distance, defined as:
\begin{equation}
TruePositiveRate: \frac{TP}{TP+FP}
\label{TruePositiveRate}
\end{equation}
\begin{equation}
FalseNegativeRate: 1 -\frac{TP}{TP+FN}
\label{FalseNegativeRate}
\end{equation}
\begin{equation}
CorrectlyPairedFraction: \frac{d(TP)-d(FP)-d(FN)}{d(TP)+d(FP)+d(FN)}
\label{CorrectlyPairedFraction}
\end{equation}
where $TP$ is the number of true positive MB pairs, $FP$ is the number of false positive MB pairs, $FN$ is the number of false negative MB pairs and $d(\cdot)$ indicates the sum of the Euclidean distance in the corresponding class of paired MBs.

\subsubsection{Evaluation for MB trajectory interpolation}
\begin{figure}[!t]
    \centerline{\includegraphics[width=\columnwidth]{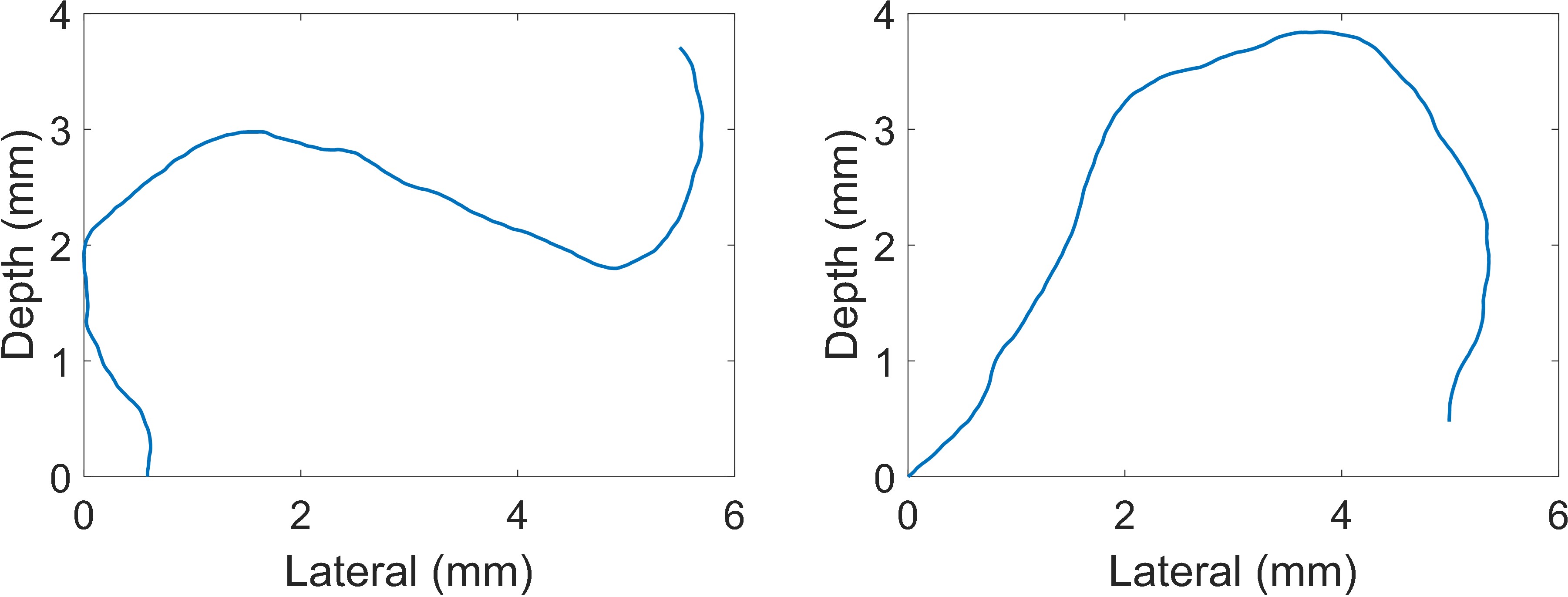}}
    \caption{Examples of microvasculature simulation datasets for interpolation.}
    \label{fig4}
\end{figure}

Six additional datasets, each containing a single vessel, were generated to test the performance of MB trajectory interpolation (Fig. \ref{fig4}). In this simulation only one MB was moved through each vessel, and captured at a frame rate of 25 Hz. The remaining parameters were kept the same as in the previous simulation. The MBs were paired and linked correctly by the proposed tracking algorithm. Two different interpolation methods, linear interpolation and acceleration-based nonlinear interpolation, were used in comparison. The linear interpolation method plotted straight trajectories between frames with MB positions, while the acceleration-based method plotted curved trajectories additionally with estimated MB velocities and accelerations, as shown in Eq. \eqref{interpolation}. The reconstruction error was calculated by a pointwise Euclidean distance between interpolated results and the ground truth.

\subsubsection{In vivo experiment}
Two ultrasound datasets of breast cancer patients were acquired at The Royal Marsden Hospital (London, UK), for a clinical trial (KORTUC Phase 2, ClinicalTrials.gov: NCT03946202) led by the The Institute of Cancer Research and The Royal Marsden NHS Foundation Trust. Ethics approval was granted by West of Scotland Research Ethics Committee (REC ref 20/WS/0019). The patients were informed by and signed on written contents. The ultrasound datasets were acquired using a Verasonics Vantage (Verasonics Inc., Kirkland, WA, USA) and a GE LE-12D probe (GE Healthcare, NY, USA) with a centre frequency of 5 MHz. 2.5 mL of SonoVue MBs (Bracco, Milan, Italy) were administered intravenously. 5 seconds of the dataset was used for the super-resolution processing. Images were acquired at a frame rate of 100 Hz using a mechanical index (MI) of 0.1. An amplitude modulation (AM) was used to  generate contrast-enhanced ultrasound (CEUS) sequences.

Tissue motion in datasets were estimated from the B-mode sequences, reconstructed from the AM pulse, using a non-rigid registration algorithm. The CEUS sequences were corrected correspondingly \cite{rueckertNonrigidRegistrationUsing1999,harputTwoStageMotionCorrection2018}. A moving-average across 11 frames around the frame of interest was subtracted from the sequence to remove remaining tissue signals. The moving-average window size was chosen after considering the frame rate, the velocity of the slowest blood flow to be captured, and the rate of tissue motion. The datasets were smoothed spatially and temporally using a Gaussian filter, and logarithmically compressed. To further reduce noises, a noise only dataset was acquired by imaging air and subtracted from the dataset of interest. The MB signal was localised by peaks in the map obtained by normalised cross-correlation with an estimated point spread function.

To evaluate the proposed tracking method at a typical imaging frame rate of clinical ultrasound system, the original dataset at 100 Hz was down-sampled 4 times temporally to 25 Hz. The acceleration-based and non-acceleration-based tracking methods were compared. To evaluate the influence of frame rate only and maintain the same data size, downsampling was conducted by 1) extracting the dataset by a time interval of 4 frames into 4 subgroups, 2) tracking MBs in each subgroup, and 3) combining all the tracking results to generate the final super-resolution map. Taking the tracking results obtained at 100 Hz as references, the tracking performance was evaluated by the consistency between the 25 Hz and 100 Hz frame rate tracking results.

\section{Results}
\label{sec:Results}
\subsection{Simulations}
\subsubsection{MB tracking}
\begin{figure}[!t]
    \centerline{\includegraphics[width=7.5cm]{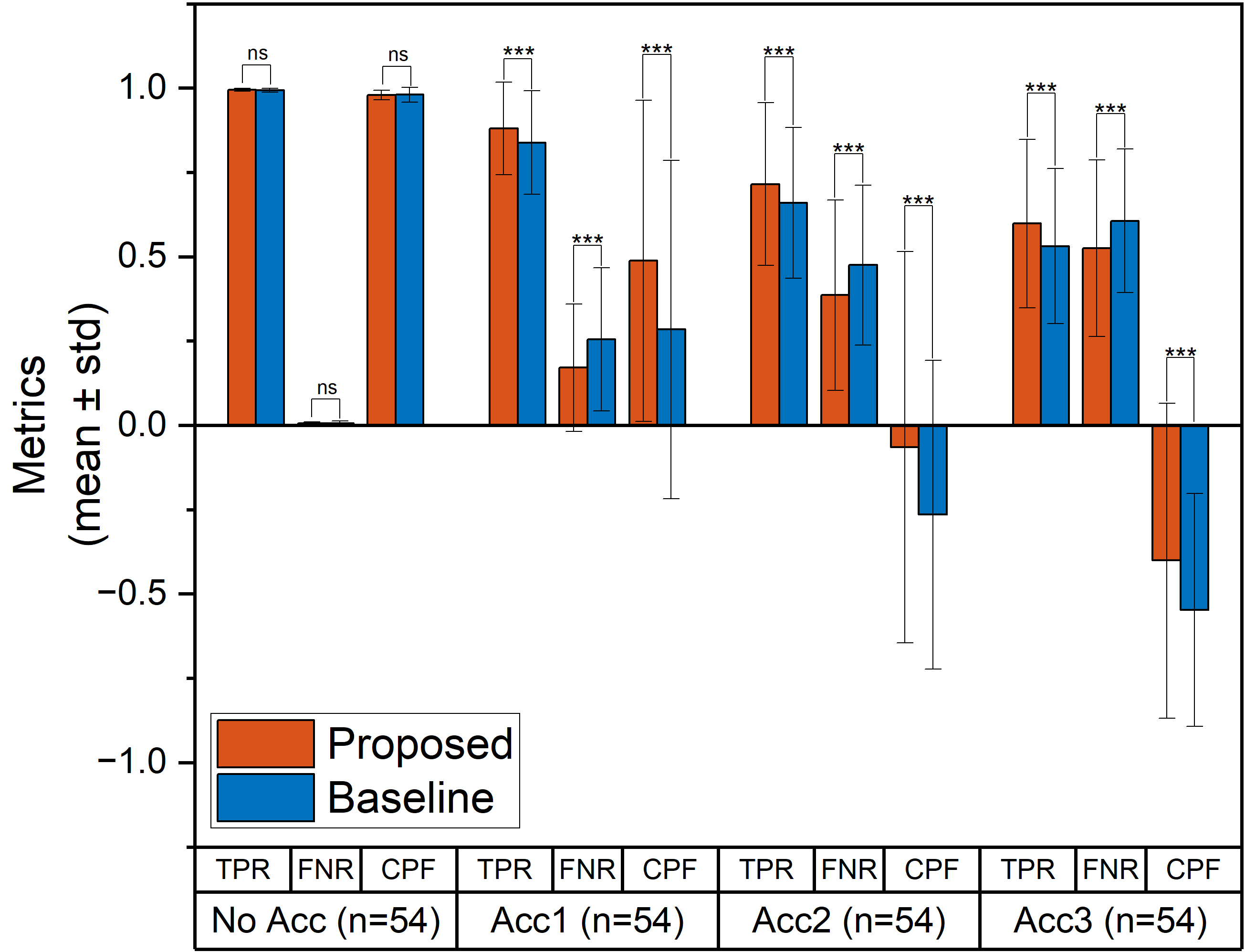}}
    \caption{Paired \emph{t}-test at different accelerations. TPR: true positive rate. FNR: false negative rate. CPF: correctly paired fraction. No Acc: no acceleration. Acc1: simulation with an acceleration of 37.5 mm/s$^2$. Acc2: 75.0 mm/s$^2$. Acc3: 112.5 mm/s$^2$. ns: no significant difference. ***: significant difference with \emph{p}$\le$0.001.}
    \label{fig_table1}
\end{figure}

\begin{figure}[!t]
    \centerline{\includegraphics[width=7.5cm]{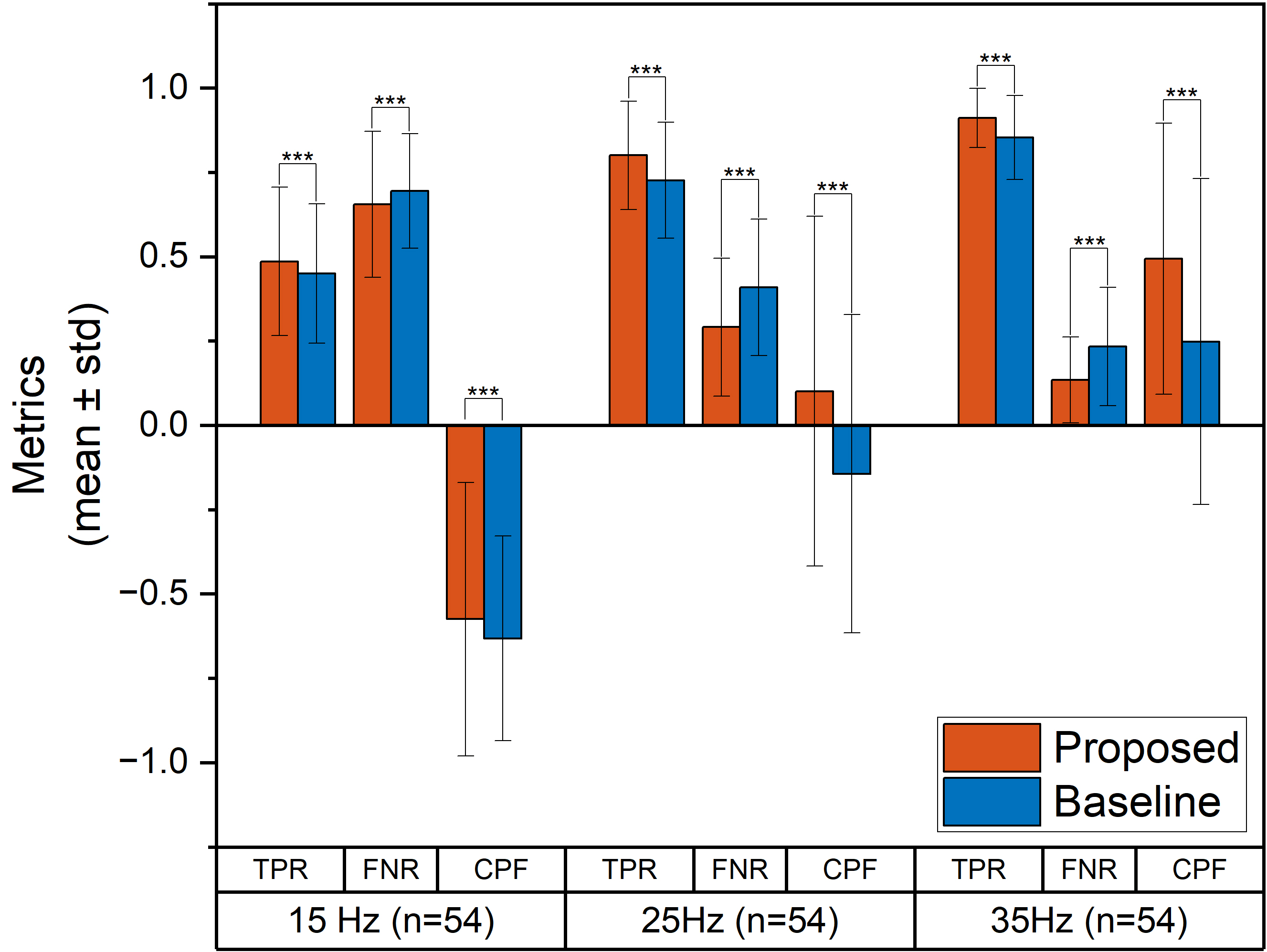}}
    \caption{Paired \emph{t}-test of different frame rates. TPR: true positive rate. FNR: false negative rate. CPF: correctly paired fraction. ***: significant difference with \emph{p}$\le$0.001.}
    \label{fig_table2}
\end{figure}

The evaluation results of MB tracking based on simulation data are shown in Fig. \ref{fig_table1} and Fig. \ref{fig_table2}.  

From the statistical analysis of MB tracking results, the proposed algorithm outperformed the baseline algorithm. When evaluating the methods with zero acceleration, there is no significant difference in false negative rate and correctly paired fraction (0.006 vs. 0.006, \emph{p}=0.115; 0.979 vs. 0.980, \emph{p}=0.42) between the results from both methods, as shown in Fig. \ref{fig_table1}. As the acceleration increased, the tracking results from the proposed method were better than the baseline, showing a significant difference in all used metrics.

To compare the tracking performance at different frame rates, we excluded those data with zero acceleration and conducted a second analysis on the tracking results at different acquisition frame rates. From the results in Fig. \ref{fig_table2}, the largest difference can be observed at a frame rate of 25 Hz, where the proposed method showed a 10.03\% (\emph{p}$<$0.001) higher true positive rate, 28.61\% (\emph{p}$<$0.001) lower false negative rate and 170.14\% (\emph{p}$<$0.001) higher correctly paired fraction.

\subsubsection{Trajectory interpolation}
\begin{figure}[!t]
    \centerline{\includegraphics[width=7.5cm]{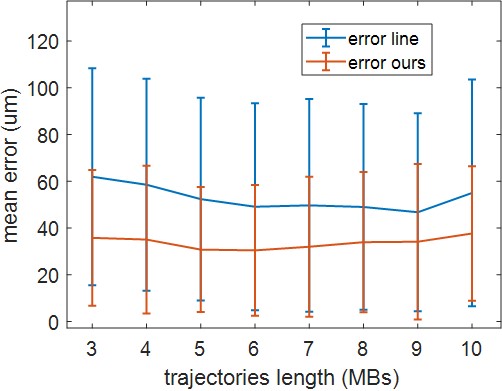}}
    \caption{Comparison of interpolation errors of two interpolation methods with different trajectories length.}
    \label{fig7}
\end{figure}

\begin{figure}[!t]
    \centerline{\includegraphics[width=7.5cm]{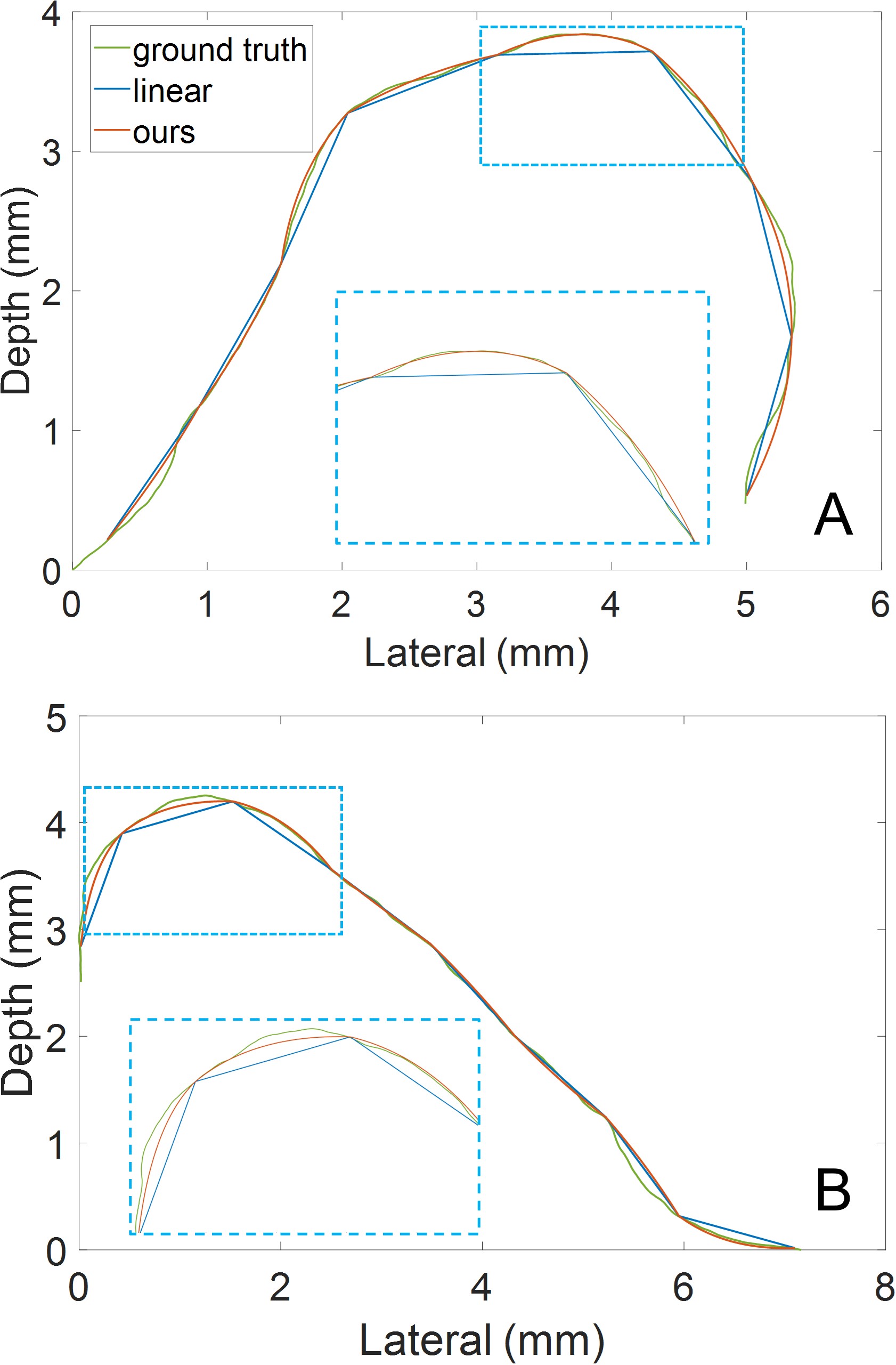}}
    \caption{Examples of MBs trajectory interpolation results via two interpolation methods. Blue boxes: ROI and zoomed-in on the difference between interpolation methods.}
    \label{fig8}
\end{figure}

The errors when interpolating between MB localisation positions to plot MB tracks were compared among the linear interpolation and acceleration-based nonlinear interpolation, and are shown in Fig. \ref{fig7}. We evaluated the interpolation error on trajectories with different numbers of linked MBs. The proposed nonlinear interpolation method showed the lowest average error of 33.71±2.52 $\mu$m. The linear interpolation showed the highest error of up to 52.80±5.29 $\mu$m on average. A visual comparison of interpolation results is shown in Fig. \ref{fig8}. Both methods showed similar results in the section where the vessels are relatively straight. When vessels were more tortuous, the nonlinear interpolation method was able to better follow the curved ground truth trajectory than the linear interpolation method.

\subsection{In Vivo Experiments}
\begin{figure*}[]
    \centerline{\includegraphics[width=18cm]{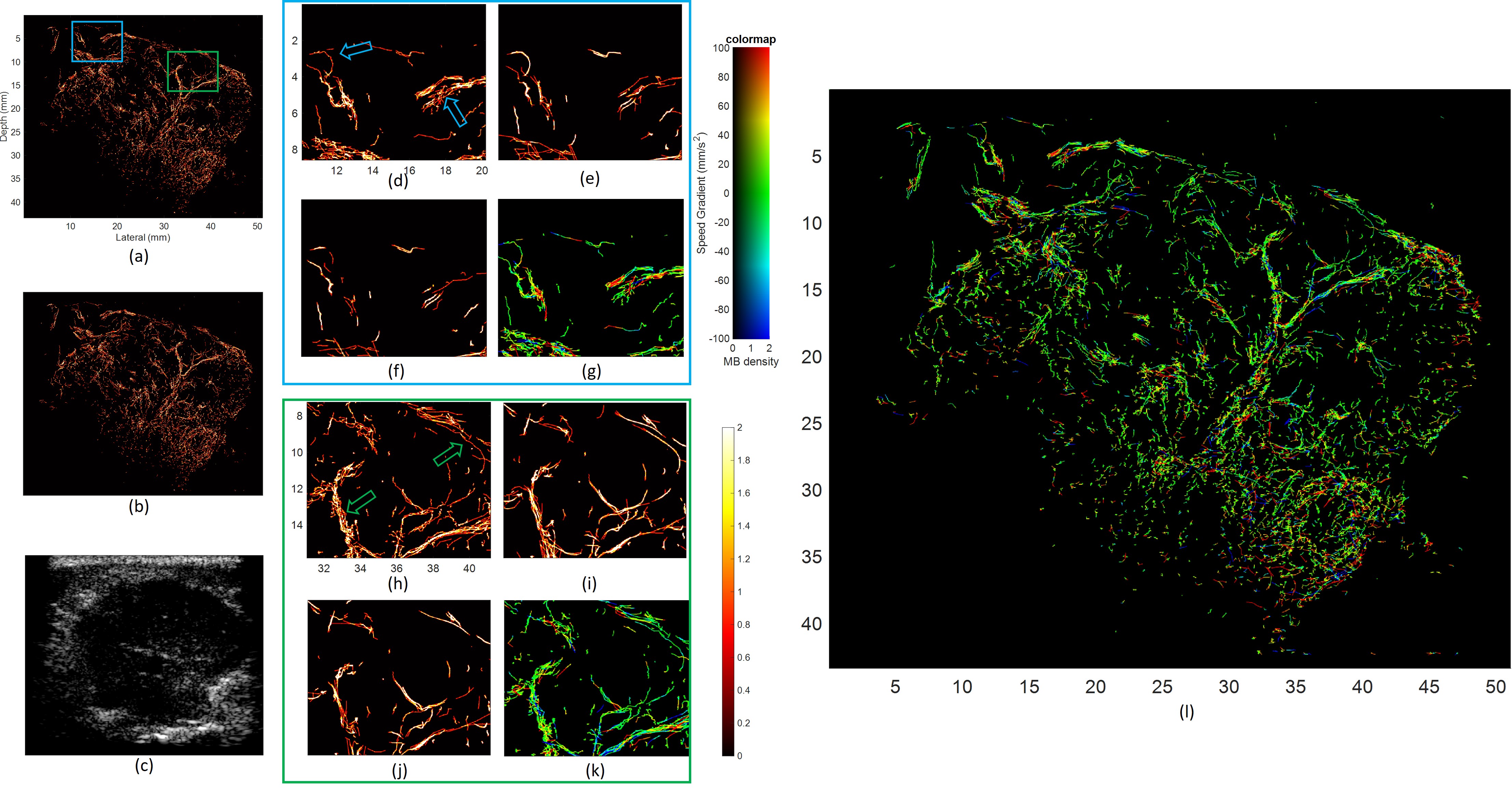}}
    \caption{Comparison between super-resolution results in the first \emph{in vivo} dataset. (a) MB tracking result in 100 Hz dataset by using the proposed method. (b) MB tracking result in 100 Hz dataset by using the baseline method. (c) One frame of B-mode images acquire from the patient. (d) Magnified blue region of interest (ROI) in (a). (e) Results from the proposed method at 25 Hz dataset in the same ROI as (d). (f) Result from the baseline method at 25 Hz dataset in the same ROI as (d). (g) Magnified spatial speed gradient in the same ROI as (d). (h-k) Result comparison of methods at the green ROI in (a). (l) A spatial speed gradient map generates from the proposed method. Blue and green arrows indicate vessels showed differences in the results from the proposed and baseline method at 25Hz. The colour bars denote the intensity of the speed gradient and the MBs' density.}
    \label{fig9}
\end{figure*}

\begin{figure*}[]
    \centerline{\includegraphics[width=18cm]{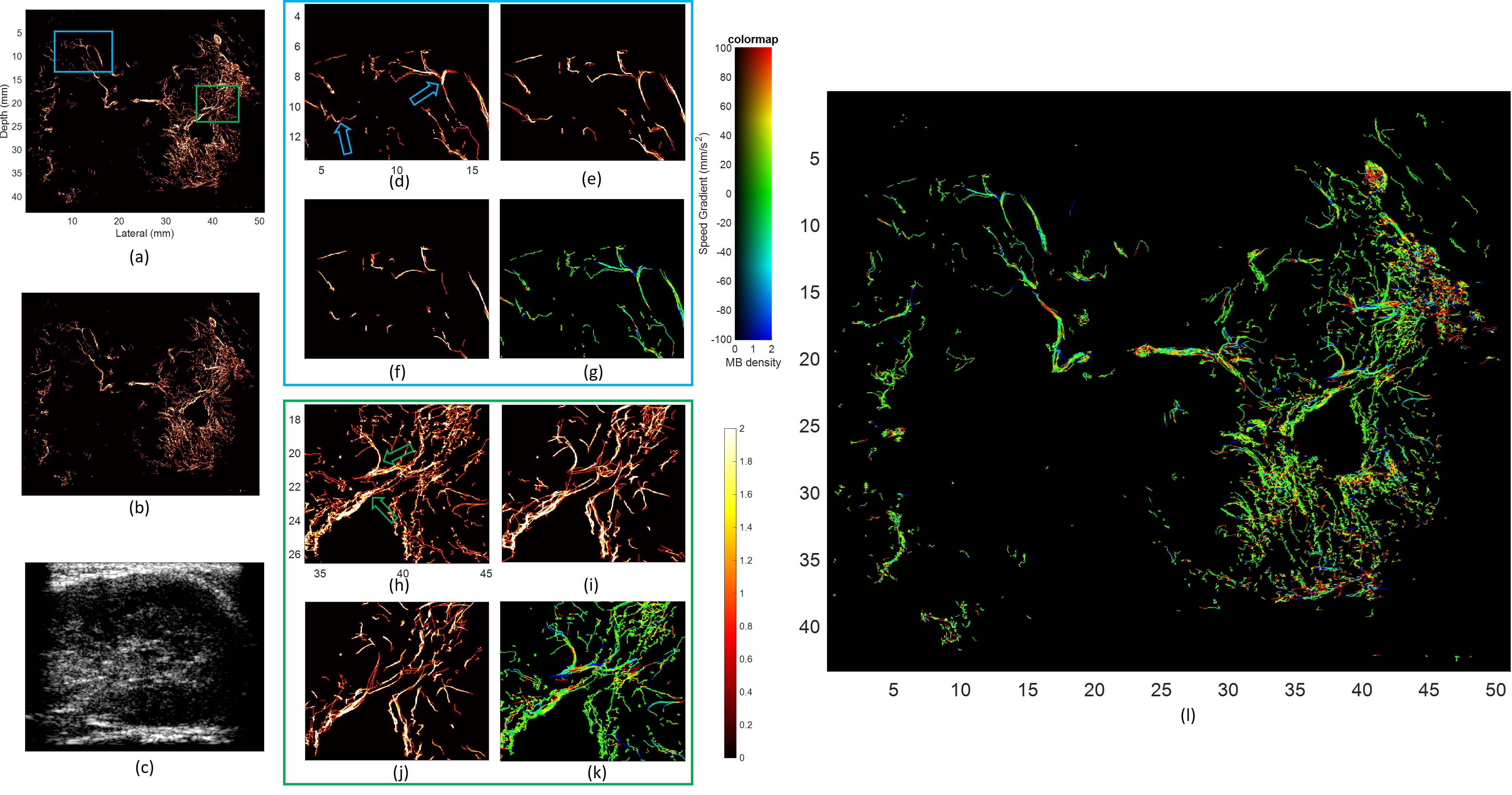}}
    \caption{Super-resolution results in the second \emph{in vivo} dataset. Captions are the same as in Fig. \ref{fig9}.}
    \label{fig10}
\end{figure*}

The comparison results are shown in Fig. \ref{fig9} and Fig. \ref{fig10}. The tracking results in the high frame rate dataset using both tracking algorithms showed little visual difference. When we down sampled the dataset 4 times, the proposed acceleration-based method outperformed the baseline method. From Fig. \ref{fig9} and Fig. \ref{fig10}, more information is kept in the 25 Hz frame rate results of the proposed method than in the baseline method. The arrows in Fig. \ref{fig9} and Fig. \ref{fig10} highlight some differences in results from the proposed and baseline methods. For the proposed method, there was a higher consistency of vessels presenting in the 25 Hz and 100 Hz results indicating a better tracking performance.

A spatial speed gradient map was generated for each of the datasets, which is readily available after the proposed acceleration-based Kalman tracking. This information is in additional to MB density and flow velocity maps and may have diagnostic value.

\section{Discussion}
\label{sec:Discussion}
\subsection{Main Findings}
In this study, we introduced an acceleration term into the Kalman-filtering-based MB tracking framework to improve MB tracking performance at low acquisition frame rates. Besides the MB tracking, incorporating acceleration also allows more accurate reconstruction of MB trajectories than linear interpolation. Results from simulation and \emph{in vivo} experiments demonstrate a significant improvement in MB tracking and vasculature reconstruction by the proposed methods. Additionally, a new kind of super-resolution map, spatial speed gradient map, is generated in this study to provide additional information potentially useful for clinical diagnoses.

The proposed algorithm was evaluated under different flow acceleration settings and acquisition frame rates, showing a consistently better tracking performance than the baseline. A tracking improvement of up to 5.58\% and 8.91\% in true positive and false negative rates, respectively, was observed when the flow acceleration was set as 75 mm/s$^2$ (Fig. \ref{fig_table1}). A significant difference was also observed on the correctly paired fraction (0.488 vs. 0.284, \emph{p}$<$0.001), which indicates more correct MB pairs and less missing MB pairs from the proposed method than the baseline method. The proposed method outperformed the baseline method under all the simulated accelerations and frame rates.

The acceleration-based motion model is less sensitive to the frame rate than the baseline motion model and gives the highest difference at a frame rate of 25 Hz. The constant acceleration assumption in the acceleration-based motion model held less when the time interval between frames increased. Thus, the improvement with the acceleration over another can be less significant when the frame rate is too low, which can be seen from the results at the frame rate of 15 Hz.

The proposed acceleration-based interpolation method had a lower reconstruction error on average when compared with the linear interpolation (33.71 $\mu$m vs. 52.8 $\mu$m, \emph{p}$<$0.001). The interpolation error of different trajectory lengths is shown in Fig. \ref{fig7}.

From the \emph{in vivo} studies, the performance between the proposed and baseline method has no significant difference when the frame rate is high. Compared with the MBs tracking results of the 100 Hz dataset, fewer MBs were tracked from the down-sampled 25 Hz dataset. Notably the proposed method can better track MBs at positions with higher spatial speed gradients. The arrows in Fig. \ref{fig9} pointed out vessel branches tracked at 100 Hz but was missed by the baseline method at 25 Hz. From the corresponding speed gradient map, a higher magnitude of acceleration (more red or blue colour) can be observed at this branch. While two curved vessels are presented with the proposed method in Fig. \ref{fig9}, the baseline method, in this case, failed to track these vessels. The proposed method benefits MBs tracking in curved vessels. In Fig. \ref{fig10}, vessels with branches were pointed, where a higher magnitude of acceleration was also presented in the speed gradient map. Again, the proposed method was able to track the vessel while the baseline method failed. The spatial speed gradient visualisation proves the proposed method's benefits of tracking the MBs with acceleration.

\subsection{Difference from Previous Works}
The nonlinear motion used for MB tracking tasks was mentioned in \cite{piepenbrockMicrobubbleTrackingNonlinear2020} to generate curved tracks. They modelled the MB movement with a constant speed and turning rate between frames. Compared with the linear motion model, their nonlinear model was a better approximation. An unscented Kalman filtering framework was used to incorporate their proposed nonlinear motion model \cite{julierUnscentedFilteringNonlinear2004}. In this paper, we proposed an acceleration motion model, approximating the MB movement as a curved motion with a changing speed. Compared with the linear motion model used in previous works, the acceleration term in the motion model avoided discontinuity in estimation of MB velocity in the tracking and is suitable for the scenario where pulsatile flow existed.

The Kalman filtering-based MB tracking framework was also used in others’ work. However, the initialisation of MB movement vectors has not been reported as far as we are aware. Inspired by the particle tracking velocimetry, we introduced a 3-frames-based MB motion state initialisation method for the first time in SRUS. A 3D graph-based algorithm was used to find the optimal initial pairing for new MBs.

The MBs’ trajectory reconstruction for super-resolution needs interpolation between linked MBs’ positions. To estimate the missing positions of MBs, linear interpolation was used in previous studies. Instead of using a fixed interpolation factor, an adaptive interpolation factor was also introduced by \cite{tangKalmanFilterBasedMicrobubble2020}. However, the hypothesis for linear interpolation that an MB kept moving with a constant velocity between frames may not hold in the case of low acquisition frame rates, tortuous vessels, and high flow speeds. In this study, we used nonlinear interpolation to reconstruct the trajectory and investigated the interpolation error between linear and nonlinear methods for the first time.

From our proposed MB tracking framework, we presented the spatial speed gradient maps for the microvasculature. The spatial gradient map may potentially indicate abnormal structural changes in the microvasculature, such as change of curvature or diameter, that results in a sudden change of MB movement.

\subsection{Limitation and Future Work}
In the MB motion state initialisation, we use a 3-frame-based method to estimate each newly appeared MB’s motion parameter. The motion model for the 3-frame initialisation assumes a linear motion for MBs between the $1^{st}$ to $2^{nd}$ frames and $2^{nd}$ to $3^{rd}$ frames. The acceleration was then estimated by the velocity change. It is worth exploring a 4-frame-based initialisation method, so the acceleration motion model can be implemented to estimate the MB’s state better. However, the higher computational cost of the 4D graph-based algorithm for 4-frame initialisation needs to be optimised in the future. The clinical application of the spatial speed gradient map from SRUS is also worth  exploring.

\section{Conclusion}
\label{sec:Conclusion}
In this paper, we introduced an acceleration-based motion model for MB tracking. A 3-frame-based motion state  initialisation method was combined with an existing Kalman tracking framework. From the evaluation of both simulation and \emph{in vivo} datasets, the proposed method is shown to improve MB tracking performance at low frame rates when there are tortuous vessels and accelerations in flows. The acceleration information can also be used for more accurate interpolation of MB trajectories between localised positions. Finally, a spatial speed gradient map is presented for the first time and could help explore the potential abnormal changes in the microvasculature.

\bibliographystyle{IEEEtran}
\bibliography{reference}

\end{document}